\documentclass[preprintnumbers,amssymb,amsmath,superscriptaddress,letterpaper,nofootinbib,showpacs]{revtex4}[10pt]

\usepackage{graphicx}
\usepackage{dcolumn}
\usepackage{bm}
\usepackage{natbib}
\usepackage{calligra}
\usepackage[T1]{fontenc}
\usepackage{egothic}
\usepackage[T1]{fontenc}
\newfont{\rsfsten}{rsfs10 scaled 1200}
\newfont{\rsfsseven}{rsfs10 scaled 1200}
\newfont{\rsfsfive}{rsfs10 scaled 1200}
\usepackage{epsfig}
\usepackage{units}
\usepackage[utf8]{inputenc}

\newcommand{\be}{\begin{equation}}
\newcommand{\ee}{\end{equation}}
\newcommand{\bea}{\begin{eqnarray}}
\newcommand{\eea}{\end{eqnarray}}

\def\lsim{\mathrel{\raise.3ex\hbox{$<$\kern-.75em\lower1ex\hbox{$\sim$}}}}
\def\gsim{\mathrel{\raise.3ex\hbox{$>$\kern-.75em\lower1ex\hbox{$\sim$}}}}

\begin{document}

\hspace*{120mm}{FERMILAB-PUB-15-477-A}
\vskip 0.2in

\title{A Predictive Analytic Model for the Solar Modulation of Cosmic Rays}

\author{Ilias Cholis}
\email{icholis1@jhu.edu}
\affiliation{Department of Physics and Astronomy, The Johns Hopkins University, Baltimore, Maryland, 21218, USA}
\affiliation{Fermi National Accelerator Laboratory, Center for Particle Astrophysics, Batavia, Illinois, 60510, USA}
\author{Dan Hooper}
\email{dhooper@fnal.gov}
\affiliation{Fermi National Accelerator Laboratory, Center for Particle Astrophysics, Batavia, Illinois, 60510, USA}
\affiliation{University of Chicago, Department of Astronomy and Astrophysics, Chicago, Illinois, 60637, USA}
\author{Tim Linden}
\email{linden.70@osu.edu}
\affiliation{University of Chicago, Kavli Institute for Cosmological Physics, Chicago, Illinois, 60637, USA}
\affiliation{Center for Cosmology and AstroParticle Physics (CCAPP) and Department of Physics, The Ohio State University Columbus, OH, 43210 }

\date{\today}

\begin{abstract}

An important factor limiting our ability to understand the production and propagation of cosmic rays pertains to the effects of heliospheric forces, commonly known as solar modulation. The solar wind is capable of generating time and charge-dependent effects on the spectrum and intensity of low energy ($\lsim$ 10~GeV) cosmic rays reaching Earth. Previous analytic treatments of solar modulation have utilized the force-field approximation, in which a simple potential is adopted whose amplitude is selected to best fit the cosmic-ray data taken over a given period of time. Making use of recently available cosmic-ray data from the \textit{Voyager 1} spacecraft, along with measurements of the heliospheric magnetic field and solar wind, we construct a time, charge and rigidity-dependent model of solar modulation that can be directly compared to data from a variety of cosmic-ray experiments. We provide a simple analytic formula that can be easily utilized in a variety of applications, allowing us to better predict the effects of solar modulation and reduce the number of free parameters involved in cosmic ray propagation models.

\end{abstract}
\pacs{96.50.S-, 96.50.sh, 95.85.Ry, 98.70.Sa}
\maketitle

\section{Introduction}
\label{sec:introduction}
Nearly all direct observations of cosmic rays (CRs) are conducted within the heliosphere of the Solar System. As a result of the turbulent solar wind and its embedded magnetic field, the CR spectrum observed at Earth can differ significantly from that in the surrounding interstellar medium. This is particularly true for CRs with kinetic energies below $\sim$10 GeV, which can be efficiently deflected and de-accelerated as they propagate through the heliosphere. The effects of solar modulation are charge-dependent and vary with time, showing a strong correlation with solar activity. 

With the deployment of instruments such as \textit{PAMELA} and \textit{AMS-02}, local CR observations have entered a high-precision era, in which statistical errors are often much smaller than the corresponding systematic uncertainties associated with CR propagation. Importantly, a myriad of free parameters in CR propagation models, such as those governing diffusive reacceleration and convection, impact primarily the same low-energy CR population that is most affected by solar modulation. Thus improvements in our understanding of solar modulation can allow for more reliable inferences of the parameters describing the injection and transport of CRs throughout the Milky Way. 



In most previous work, members of the CR and particle physics communities have employed the ``force-field'' approximation to describe the effects of solar modulation on the observed CR flux. In this approach, the kinetic energy  ($E_{kin}$) of each charged particle is simply reduced by a quantity $\mid Z \mid e\Phi$, where $\Phi$ is known as the modulation potential, which is generally found to be on the order of 0.1-1 GV. The effect of the modulation potential on the CR spectrum can be written as follows \cite{1968ApJ...154.1011G}:
\begin{equation}
\frac{dN^{\oplus}}{dE_{kin}} (E_{kin}) = \frac{(E_{kin}+m)^{2} -m^{2}}{(E_{kin}+m+\mid Z\mid e \Phi)^{2} -m^2} \,\, \frac{dN^{\rm ISM}}{dE_{kin}} (E_{kin}+\mid Z\mid e \Phi),
\end{equation}
where $E_{kin}$ is the observed kinetic energy and the subscripts ``ISM'' and ``$\oplus$'' denote values in the interstellar medium and at the location of Earth, respectively. Also $\mid Z \mid e$  refers to the absolute charge of CRs.
To address the time variability of solar modulation, one typically adopts a value for $\Phi$ that provides the best fit to the data collected over a given period of time. It is also common for studies to adopt different values of $\Phi$ for positively and negatively charged CRs, allowing for the possibility of charge-sign dependent effects. Yet, because of its simplicity, there are considerable weaknesses associated with this approach. In particular, the force-field approximation \cite{1968ApJ...154.1011G} does not allow for any rigidity dependent effects (rigidity : $R \equiv p/q$, $p$ is the CR momentum and $q=Ze$ the charge), and fits to a given cosmic-ray dataset often find significant degeneracies between the modulation potential and the parameters describing Galactic CR injection and transport. Furthermore, this approach cannot predict the behavior of the modulation potential with time, and thus cannot be used to compare datasets taken over different periods.

A second approach employed in recent years to account for the effects of solar modulation involves the use of highly sophisticated particle propagation codes to model the physical processes of three-dimensional diffusion, particle drifts, convection and adiabatic energy losses~\cite{2012Ap&SS.339..223S, 2013PhRvL.110h1101M} \footnote{Some recent results using implementations of those codes have been shown in \cite{Bisschoff:2015qed, Evoli:2015vaa, Gaggero:2013nfa}.}. This approach is physically well motivated, and in many ways provides an effective technique for calculating the impact of solar modulation. These particle propagation codes have several weaknesses, however, which currently limit their utility. Firstly, they include large numbers of free-parameters which must be scanned over in parallel with parameters associated with CR injection and propagation. This makes such approaches computationally intensive, preventing their usage in most CR propagation parameter space scans. Secondly, the Solar System propagation codes of this nature are not currently publicly available, limiting their utility to the broader CR community.

In this paper, we approach this problem by constructing an empirical model for the modulation potential that is time, charge, and rigidity dependent. We encapsulate this model in an analytic formula that is nearly as simple to implement as the force-field approximation, but that has several significant advantages. In particular, by making use of solar physics observables that are independent of CR propagation parameters, we are able to predict the solar modulation potential over different periods of time, allowing us to compare the results of multiple CR experiments, as opposed to treating each experiment's modulation parameter as an independent nuisance parameter.
%
%
%
%
There are three key factors that have made it possible for us to model solar modulation in this way:

\begin{itemize}
\item{Several well measured solar observables are known to correlate with the solar modulation potential, including the magnitude of the solar magnetic field, the bulk velocity of the solar wind, and the tilt angle of the heliospheric current sheet. 
} 
\item{The vast CR datasets provided by the \textit{PAMELA}~\citep{Adriani:2008zq, Adriani:2010rc, Adriani:2012paa} and \textit{AMS-02}~\cite{AMS02,Aguilar:2015ooa,Aguilar:2014fea,Aguilar:2014mma} experiments have made it possible to measure variations in the local CR spectrum over relatively short timescales with high statistical precision.}
\item{In the summer of 2012, the\textit{Voyager 1} spacecraft passed through the heliopause, where it directly measured the CR spectrum unaffected by the influence of the solar wind for the first time~\citep{2013Sci...341..150S}.}
\end{itemize}


Our model for solar modulation employs three quantities that are well-studied in solar physics: the polarity of the solar magnetic field, the magnitude of the heliospheric magnetic field (HMF) at the position of Earth~\cite{ACESite}, and the tilt angle of the heliospheric current sheet~\cite{WSOSite}. We take advantage of the fact that these solar observables, and the CR modulation potential, evolve on monthly to yearly timescales, while the CR flux in the local interstellar medium is effectively in steady state.
%
%
To constrain the free parameters in our model, we make use of measurements of the CR proton flux, antiproton flux, and the ratio of boron-to-carbon nuclei, as reported by BESS \cite{1994AdSpR..14...75Y}, BESS Polar~\cite{Abe:2015mga}, IMAX~\cite{2000ApJ...533..281M}, CAPRICE \cite{Boezio:2002ha}, \textit{PAMELA}~\citep{Adriani:2008zq, Adriani:2010rc, Adriani:2012paa}, \textit{AMS-01} \cite{Alcaraz:2000vp}, \textit{AMS-02} \cite{AMSsite}, and \textit{Voyager 1}~\citep{2013Sci...341..150S, Potgieter:2013mcc} 
\footnote{\cite{Maurin:2013lwa} is also a useful database for the CR data by various experiments.}.
Ultimately, after considering several physically motivated functional forms, we arrive at the following analytic expression for the solar modulation potential:
\begin{eqnarray}
\Phi(R,t) = \phi_{0} \, \bigg( \frac{|B_{\rm tot}(t)|}{4\, {\rm nT}}\bigg) + \phi_{1} \, H(-qA(t))\, \bigg( \frac{|B_{\rm tot}(t)|}{4\,  {\rm nT}}\bigg) \, \bigg(\frac{1+(R/R_0)^2}{\beta (R/R_{0})^3}\bigg) \, \bigg( \frac{\alpha(t)}{\pi/2} \bigg)^{4},
\label{eq:ModPot_Intro}
\end{eqnarray}
%
%
where $|B_{\rm tot}|$ and $A$ are the strength and polarity of the HMF (as measured at Earth), and $\alpha$ is the tilt angle of the heliospheric current sheet. These quantities are treated as time-dependent inputs, independent of CR observables. $R$, $\beta$, and $q$ are the rigidity, velocity, and charge of the CR, respectively. $H$ is the Heaviside step function and is equal to zero or unity depending on the product of the charge of the CR and the polarity of the HMF. By fitting our analytic formula to a variety of available CR data, we determine the best-fit
values of the reference rigidity, $R_0 = 0.5$ GV, and the normalization
factors, $\phi_0 \simeq 0.35$ GV and $\phi_1 \simeq 3.9$ GV.

In the remainder of this paper, we will discuss the physical basis for this formula, describe its robustness to model assumptions, and its utility to ongoing studies of CR propagation. Specifically, in Section~\ref{sec:PropThroughSolSyst}, we review and discuss the physics of CR propagation through the heliosphere. In Section~\ref{sec:ISM_Uncer}, we utilize the CR propagation code {\tt Galprop}, selecting several sets of parameters to demonstrate that our results are robust to such variations. In Section~\ref{sec:set_up}, we directly calculate the free parameters in our theoretically driven solar modulation model through a comparison with various CR datasets. Finally, in Section~\ref{sec:Conclusions}, we summarize our results and conclusions.

\section{The Propagation of Cosmic Rays Through The Solar System} 
\label{sec:PropThroughSolSyst}

In this section, we describe the major factors regulating the transport of CRs through the Solar System. The solar wind consists of a stream of $\sim$1-10 keV electrons, protons, and helium nuclei that is projected from the upper atmosphere of the Sun. The intensity and spectrum of this emission varies with time, and across the solar surface. As the solar wind flows outward from the Sun, it fills a volume of space known as the heliosphere, which is bounded at the heliopause where the pressure of the solar wind is balanced by the interstellar medium. The geometry of the heliosphere is believed to be highly asymmetric, resembling a bubble with a long cometary tail. The distance to the heliopause varies with time, but is typically on the order of $\sim$\,$100$~AU. The first direct observation of the heliopause was made on August 25, 2012, when the \textit{Voyager 1} spacecraft measured the local plasma density to suddenly increase by a factor of $\sim$40.
%
%
Within the heliosphere resides a more spherical boundary called the termination shock, centered on the sun with a radius of $\sim$80-100 AU. The termination shock represents the point at which the velocity of the solar wind falls below the sound speed of the interstellar medium ($\sim$100 km/s). \textit{Voyager 1} and \textit{2} have each crossed the termination shock, in 2004 and 2007,  and at distances of 94 and 84 AU, respectively \cite{2005Sci...309.2017S, 2008Natur.454...71S}.


Among other phenomena, the solar wind is responsible for the HMF. The HMF exhibits a spiral structure on large scales, with an average magnitude that falls with the square of the distance to the Sun, and which is typically $\sim$4-8 nT at Earth (averaged over month-long timescales). Measurements of the HMF show significant time variation, both at the solar surface and in near-Earth orbit. The most readily apparent feature of the HMF is its $\sim$22~year cycle, which includes a reversal in polarity, $A$, every 11~years.\footnote{Negative (positive) polarity of the HMF refers to the case in which the coronal magnetic field lines point inward (outward) from the north pole of the Sun.} The last two HMF polarity reversals occurred between October 1999 and June 2000 (from $A>0$ to $A<0$) and between October 2012 and June 2013 (from $A<0$ to $A>0$). Observations by the \textit{Voyager} probes have observed field reversals in the outer Solar System that are temporally correlated to those observed near the Earth and Sun, demonstrating that the time variation of the HMF is a global phenomenon.


The propagation of CRs through the HMF can be described by:
\begin{eqnarray}
 \frac{\partial f}{\partial t}&=& -(\vec{V} + \langle \vec{v}_D \rangle) \nabla f + \nabla (\hat{D}\nabla f) + \frac{1}{3} (\nabla \vec{V}) \frac{\partial f}{\partial \ln p} +J_{\rm source},
\label{eq:fullPropEq}
\end{eqnarray}
where $f$ is the CR phase space density, $\vec{V}$ is the solar wind velocity, $\langle \vec{v}_D \rangle$
is the average drift velocity, $\hat{D}$ is the diffusion tensor, and $J_{\rm source}$ is a source term associated with CRs that are produced within the heliosphere, such as Jovian electrons or pick-up ions~\cite{Strauss:2012baa}. 
Equation~\ref{eq:fullPropEq} accounts for five physical phenomena: convection and drift (first term on the right hand side), diffusion (second term), adiabatic energy losses (third term), and CR sources (final term). For the range of magnetic field strengths observed at Earth, the source term can be safely ignored for CRs with $R \gsim 0.5$ GV. Additionally, although the reacceleration of CRs at the heliosheath can be important for CRs with $R \lsim 0.2$ GV (corresponding to $E_{\rm kin} \lsim 0.02$ GeV for protons), adiabatic energy losses dominate for higher energy CRs. 



Gradients and curvatures in the HMF cause CRs to drift, with an average velocity given by~\cite{2000JGR...10527447B,Strauss:2012baa}:
\begin{eqnarray}
\langle \vec{v}_D \rangle &=& \frac{q v}{3} \,\,\, \nabla  \times (\lambda_d \, \hat{e}_B),
\label{drift1}
\end{eqnarray}
where $q$ and $v$ are the charge and speed of the CR, $\hat{e}_B$ is the unit vector in the direction of the magnetic field, and $\lambda_d$ is the drift scale, given by:
\begin{eqnarray}
\lambda_d = r_{\rm Larmor} \, \frac{(R/R_0)^2}{1+(R/R_0)^2},
\label{drift2} 
\end{eqnarray}
where $r_{\rm Larmor}=p/|q|B$ is the particle's Larmor Radius. At low rigidities, the Larmor Radius of a CR is much smaller than the curvature of the HMF, and particle trajectories follow the local magnetic field structure, suppressing the drift velocity (as well as any diffusion perpendicular to the HMF lines). In contrast, at high rigidities CRs are not affected by the small-scale structure of the HMF field lines, but instead probe the average HMF structure and intensity, $\lambda_d \sim r_{\rm Larmor}$~\cite{2000JGR...10527447B}. The reference rigidity, $R_0 \sim \mathcal{O}(1)$ GV, is a free parameter that sets the scale at which the transition between these two limiting regimes occurs.


Combining Eqns.~\ref{drift1} and~\ref{drift2}, we find the timescale for CR drift to be proportional to the following:
\begin{eqnarray}
\tau_{D} \propto \frac{1}{|\langle \vec{v}_D \rangle|} \propto B(t) \, \frac{1+(R/R_0)^2}{\beta \, (R/R_0)^3},
\label{drift3}
\end{eqnarray}
where $\beta=v/c$. 
%
The drift timescale is thus expected to have the same time-dependence as the HMF, allowing us to differentiate the effects of solar modulation from those associated with propagation through the interstellar medium. 


\begin{figure}[!t]
\includegraphics[width=5.40in,angle=0]{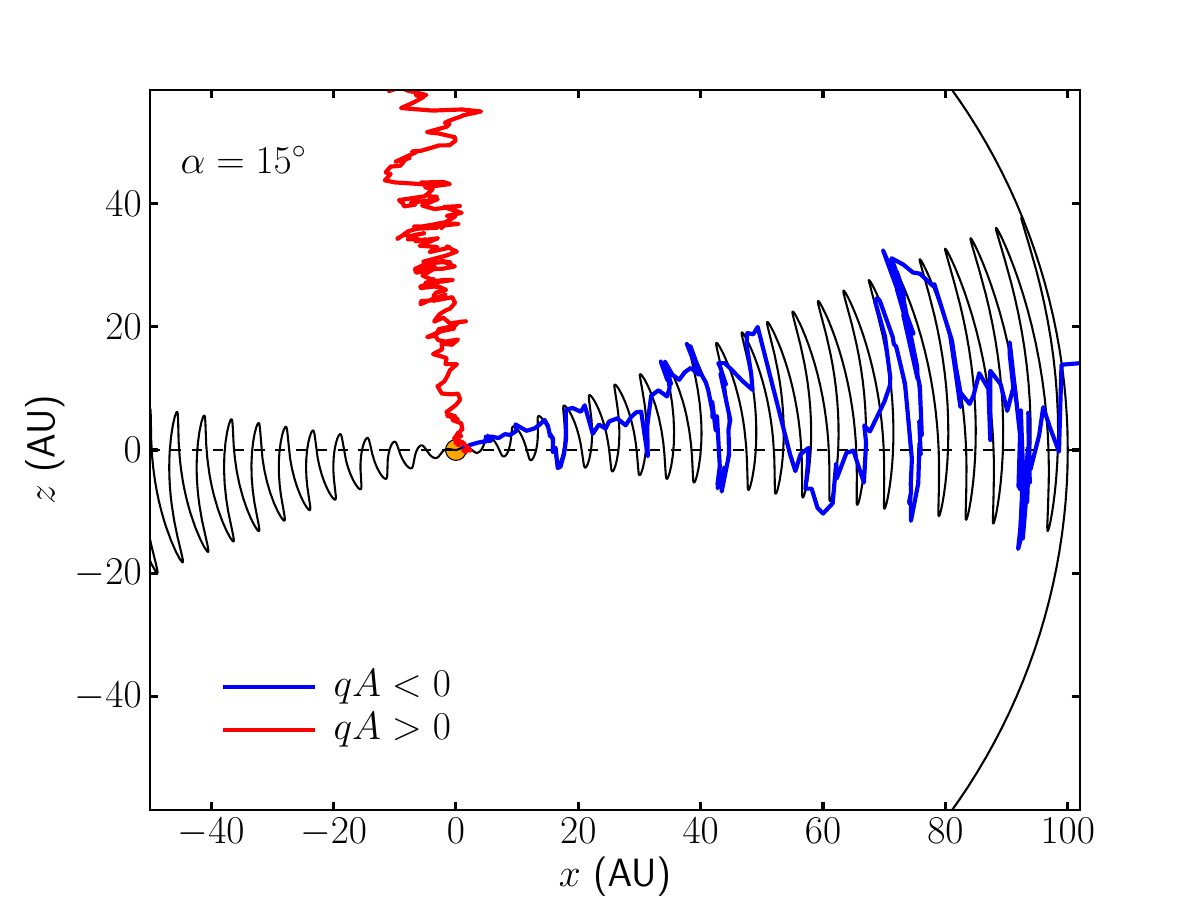}
\caption{The schematic depiction of CR propagation through the Solar System (viewed edge-on). CRs that reach the Earth follow very different trajectories, depending on the polarity of the heliospheric magnetic field. In the negative polarity state ($A<0$), positively charged CRs drift along the heliospheric current sheet (shown as a periodic solid line, for the case of a tilt angle of $\alpha=15^{\circ}$), and move across layers of this sheet via diffusion. In contrast, positively charged particles diffuse more directly and efficiently during periods of positive polarity ($A>0$). As a consequence, propagation times and corresponding energy losses can vary significantly depending on the period of the solar cycle, and on the sign of the charge of the CR. The curved line represents the boundary of the Solar System, corresponding to the region near the heliopause and/or termination shock.}
\label{schematic}
\end{figure}



During periods of positive polarity ($A>0$), a proton that originates from the polar regions of the heliopause can rather directly and effectively propagate to the location of Earth, suffering only modest energy losses. In contrast, during periods of negative polarity ($A<0$), CR protons travel toward the inner Solar System largely through regions near the plane of the Solar System, where their movement is dominated by drift along the heliospheric current sheet.

The heliospheric current sheet is the surface across which the polarity of the HMF changes. As a result of the Sun's rotation and its spiral-shaped magnetic field, this sheet is wavy, rippling periodically above and below the plane perpendicular to the Sun's rotational axis (which is inclined $7.25^{\circ}$ relative to the ecliptic). An electrical current on the order of $10^{-10}$ A$/$m$^2$ flows along the sheet. The inclination of the heliospheric current sheet with respect to the solar rotation plane varies with time, and is described by the tilt angle, $\alpha(t)$. Any particle which travels from the heliopause to the Earth along the heliospheric current sheet must propagate over an extremely long distance, especially during periods with large $\alpha$.

For the case of propagation from the poles (occurring largely during periods with $qA>0$), CR propagation is expected to be nearly independent of $\alpha$. For propagation through the heliospheric current sheet ($qA<0$), however, the energy losses incurred should increase with increasing tilt angle. Although it is difficult to predict the detailed functional relationship between the modulation potential and $\alpha$, we can constrain this function with observations. Additionally, we note that CR propagation should be independent of $\alpha$ in the high-rigidity limit, for which particle diffusion dominates over propagation along local magnetic field lines. 


In Figure~\ref{schematic}, we present a schematic depiction of CR propagation through the heliosphere. The red line represents the trajectory of a positively charged CR during a period of positive polarity (or, alternatively, a negatively charged CR during a period of negative polarity). In this case, particles propagate efficiently to Earth, suffering only modest energy losses. In contrast, when the particle charge and solar polarity are opposite (blue line), CRs propagate from the heliopause to Earth along the heliospheric current sheet, and suffer significant energy losses during their lengthy trajectory. The geometry of the heliospheric current sheet shown in Figure~\ref{schematic} corresponds to a tilt angle of $\alpha=15^{\circ}$ (i.e. the angular width of the heliospheric current sheet, viewed edge-on, is $2\alpha=30^{\circ}$). For small values of the tilt angle, propagation becomes more direct, resembling the trajectories shown for $qA>0$. For very large tilt angles, the path length along the current sheet becomes untenably long, and perpendicular diffusion begins to dominate propagation.

For both $qA>0$ and $qA<0$, the CR energy losses due to solar modulation are adiabatic, and are expected to be proportional to the time taken to travel between the heliopause and Earth. For CRs traveling from the poles with a direct path length, the solar modulation potential is then directly related to the amplitude of the HMF. On the other hand, for CRs traveling through the current sheet, there is a second term that scales with the drift time defined in Equation~\ref{drift3} and additionally depends on the tilt angle, $\alpha$. We note that the separation of the solar modulation potential into rigidity dependent and independent terms was previously suggested in Ref.~\cite{2000JGR...10527447B}.



Taking into account the considerations described in this section, we adopt the following physically motivated parameterization for the solar modulation potential: 
\begin{eqnarray}
\Phi(R,t) = \phi_{0} \, g(|B_{\rm tot}(t)|) + \phi_{1} \, H(-qA)\, g(|B_{\rm tot}(t)|) \, f(\alpha(t)) \, \bigg(\frac{1+(R/R_0)^2}{\beta (R/R_{0})^3}\bigg), 
\label{eq:main}
\end{eqnarray}
where again, $|B_{\rm tot}|$ is the strength of the HMF as measured at Earth, $\alpha$ is the heliospheric tilt angle, and $A$ is the polarity of the magnetic field. The polarity, along with the CR's charge, determines the value of the Heaviside step function, $H$. $\phi_{0}$, $\phi_{1}$ and $R_0$ are free parameters which we will fit to the data.  $g(|B_{\rm tot}|)$ and $f(\alpha)$ are functions of the magnetic field and tilt angle, respectively, whose forms we will empirically constrain in Section~\ref{sec:set_up}. Although this expression is quite general, it relies on some simplifying assumptions. Perhaps most significantly, it assumes that the dependance of the modulation potential on the strength of the HMF is the same for $qA>0$ and $qA<0$. We also note that we expect this Equation to be applicable for CRs with rigidities $R \gsim R_{0}$. For $R \ll R_{0}$, drift becomes highly suppressed and propagation relies again only on diffusion. In what follows, we will test the validity of this parameterization, and use the available CR data to constrain the value of each free parameter. As we will demonstrate, for appropriate choices of $g(|B_{\rm tot}|)$ and $f(\alpha)$, this equation provides a good description for the solar modulation potential, including its variation with time, rigidity and charge.

%




\section{The Cosmic-Ray Spectrum in the Interstellar Medium}
\label{sec:ISM_Uncer}

To model the injection and propagation of CRs through the interstellar medium of the Milky Way, we make use of the publicly available \emph{Galprop} v54 1.984 code~\cite{GALPROPSite, Strong:2015zva, NEWGALPROP}, which numerically solves the following transport equation:
\begin{eqnarray}
\frac{\partial \psi(r,p,t)}{\partial t} =q(r,p,t)+\vec{\nabla} \cdot
(D_{xx}\vec{\nabla}\psi)+\frac{\partial}{\partial p}\Big[p^2D_{pp}\frac{\partial}{\partial p}
(\frac{\psi}{p^2})\Big]+
\frac{\partial}{\partial p}\Big[\frac{p}{3}(\vec{\nabla} \cdot \vec{V})
\psi\Big],
\label{eq:GalacticProp}
\end{eqnarray}
where $\psi(r,p,t) = 4\pi p^{2}f(r,p,t)$, with $f(r,p,t)$ the CR phase space density, $D_{xx}(r)$ is the spatial diffusion tensor and $D_{pp}(r)$  the diffusion tensor in momentum space. Convection perpendicular to the Galactic Disk is described by the rightmost term. For our calculations, we (safely) ignore energy losses and secondary production due to inelastic $pp$ collisions. For the source term, $q(r,p,t)$, we assume a spatial distribution following that of supernova remnants, with a broken power-law spectrum:
\begin{eqnarray}
\frac{dN_{p}}{dR} \propto \begin{cases} R^{-\alpha_{1}}, \,\,\,\,\,\,\, R<R_{\rm br}, \\ 
R^{-\alpha_{2}}, \,\,\,\,\,\,\,  R<R_{\rm br}.
\end{cases}
\label{eq:Proton_Spectr}
\end{eqnarray}

\emph{Galprop} assumes isotropic and homogeneous diffusion, described by:
\begin{equation}
    D_{xx}(R) = D_{0} \left(\frac{R}{3 \, GV}\right)^{\delta}\,,
    \label{eqn:Diffusion}
\end{equation}
where $D_{0}$ and $\delta$ are the diffusion coefficient and diffusion index. Convection is assumed to have a constant gradient, $dv_{c}/dz$, perpendicular to the disk, while reacceleration is described by:
\begin{equation}
    D_{pp}(R) = \frac{4}{3 \delta (2-\delta)(4-\delta)(2+\delta)} \frac{R^{2} v_{A}^{2}}{D_{0}(R)},
    \label{eqn:Reacceleration}
\end{equation}                                                                                                                                 
where $v_{A}$ is the Alfv$\acute{\textrm{e}}$n speed \cite{1994ApJ...431..705S}. 

\begin{figure*}
\begin{centering}
\includegraphics[width=5.30in,angle=0]{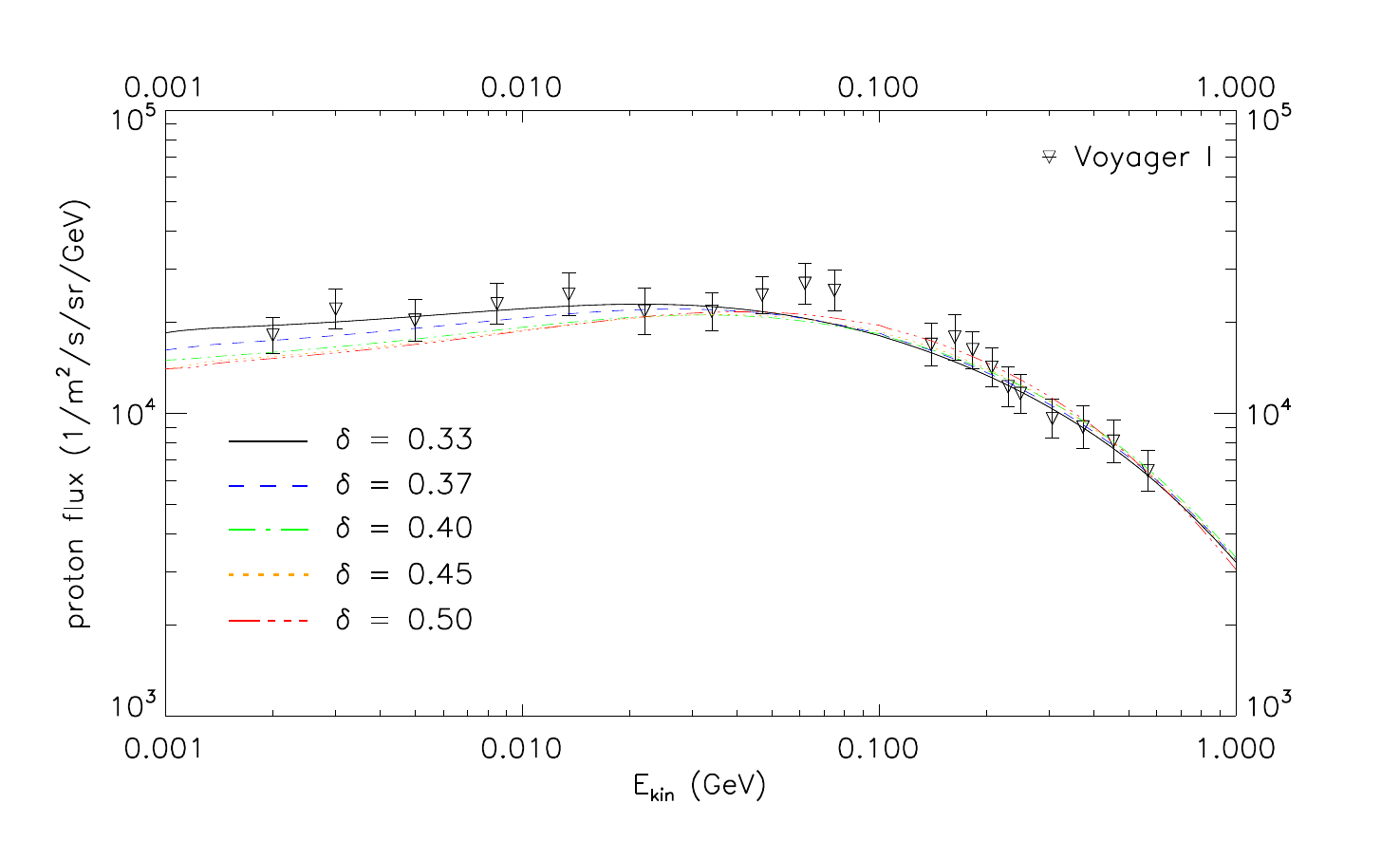}
\end{centering}
\caption{The cosmic ray proton spectrum predicted for the five Galactic cosmic-ray models described in Table~\ref{tab:fitResults} compared to the measurement of \textit{Voyager 1} \cite{Potgieter:2013mcc}.}
\label{fig:voyagerfit}
\end{figure*}

In Table~\ref{tab:fitResults}, we show the parameters for the five Galactic CR models that we will utilize throughout this study. These models cover the theoretically motivated range of $0.33 \leq \delta \leq 0.5$, spanning predictions from Komogorov to Kraichnian turbulence. In all cases we need the presence of reacceleration with Alfv$\acute{\textrm{e}}$n speeds of 23-30 km/s (see though also \cite{Drury:2015zma} for a discussion regarding the lack of diffusive reacceleration when fitting the B/C ratio data). 
In Figure~\ref{fig:voyagerfit}, we compare the CR proton spectrum predicted in each of these five models with that measured by the \textit{Voyager 1} spacecraft. These measurements are particularly powerful, as they represent the first direct measurement of the CR spectrum in the interstellar medium, before CRs experience the effects of solar modulation. The predictions of these five models are each in good agreement with this measurement.\footnote{We note that the proton flux observed by \textit{Voyager 1} below $\sim$$100$ MeV may be impacted by CR reacceleration taking place between the heliopause and the outer heliosheath~\cite{Potgieter:2013pdj}.} In Appendix~\ref{app2}, we show that these models each also provide good fits to the measured CR proton spectrum and boron-to-carbon ratio, as measured at Earth, after applying the model presented in this paper to account for the effects of solar modulation.





\begin{table*}[t]
    \begin{tabular}{ccccccccc}
         \hline
           Model & $\delta$ & $z_{L} (kpc)$ & $D_{0} \times 10^{28}$ (cm$^2$/s) & $v_{A}$ (km/s) & $dv_{c}/dz$ (km/s/kpc) & $\alpha_{1}$ & $\alpha_{2}$ & $R_{\rm br}$ (GV)\\
            \hline \hline
            A  & 0.33 & 6.0 & 6.50 & 30.0 & 0.0 & 1.95 & 2.41 & 14.3 \\
            B &  0.37 & 5.5 & 5.50 & 30.0 & 2.5 & 1.89 & 2.38 & 11.7 \\
            C &  0.40 & 5.6 & 4.85 & 24.0 & 1.0 & 1.88 & 2.38 & 11.7 \\
            D &  0.45 & 5.7 & 3.90 & 25.7 & 6.0 & 1.88 & 2.36 & 11.7 \\      
            E &  0.50 & 6.0 & 3.10 & 23.0 & 9.0 & 1.88 & 2.45 & 11.7 \\
            \hline \hline 
        \end{tabular}
    \caption{The parameters of the five models used in this study to describe the injection and Galactic propagation of cosmic rays. See text for details.}
    \label{tab:fitResults}
\end{table*}

\section{Combining Solar and Cosmic-Ray Data}
\label{sec:set_up}

\begin{figure*}
\begin{centering}
\includegraphics[width=7.2in,angle=0]{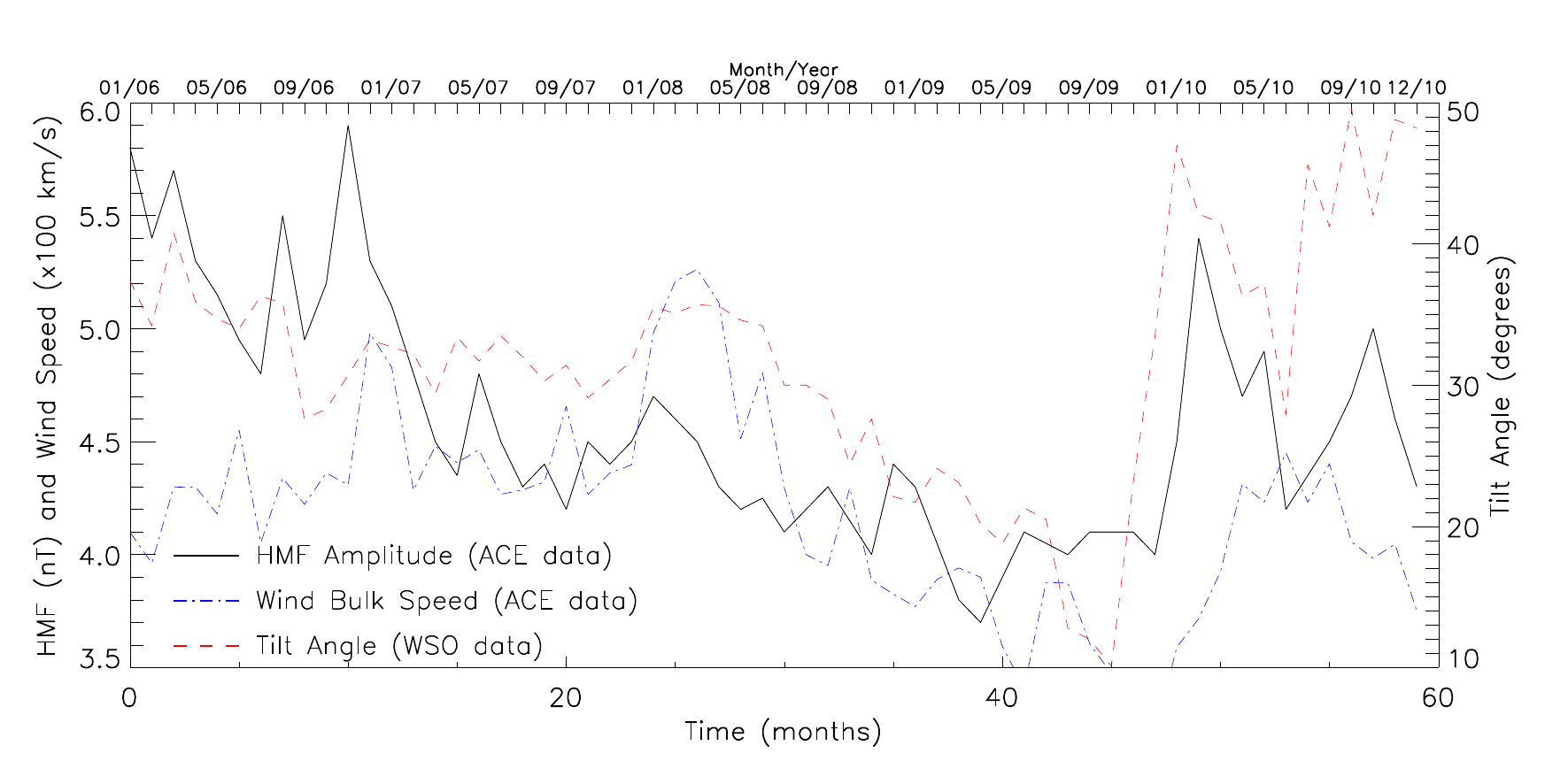}
\end{centering}
\caption{The amplitude of the heliospheric magnetic field (HMF) as measured at 1 AU~\cite{1998SSRv...86..613S, ACESite} (black solid), the solar wind bulk speed as measured at 1 AU \cite{ACESite} (blue dot-dashed), and the inferred tilt angle of the heliospheric current sheet \cite{WSOSite} (red dashed), as a function of time. Monthly averages of each quantity are shown over a five year period spanning 2006 to 2010.}
\label{fig:HMFvstime}
\end{figure*}

The local properties of the solar magnetic field have been extensively studied. In Figure~\ref{fig:HMFvstime}, we plot the observed values of the HMF amplitude at the Earth's position, $|B_{\rm tot}|$, as well as those of the bulk solar wind speed at Earth, and the HMF tilt angle, $\alpha$. The first two of these quantities were directly measured by the Advanced Composition Explorer (\textit{ACE}) Magnetometer and the Solar Wind Electron Proton Alpha Monitor (SWEPAM), respectively \cite{1998SSRv...86..613S, ACESite,SWEPAMsite}, while the value of the tilt angle has been derived from a model utilizing publicly available data from the Wilcox Solar Observatory~\cite{WSOSite} (see also, Ref.~\cite{Ng:2015gya}). We show this data over a five year period between January 2006 and December 2010, roughly corresponding to the era of \textit{PAMELA} data collection.


As expected, we find a significant degree of correlation between these three quantities. Each, for example, experiences a minimum in the summer of 2009. 
We note that while the amplitude of the local HMF varies by approximately 50$\%$ (3.7 nT --- 5.9 nT) in the monthly average, significantly larger day-to-day variations are recorded. These high-frequency variations, however, are unlikely to be correlated over the entire heliosphere and will effectively be averaged out over the 100--300 day propagation time of $\sim$100~MeV CRs through the heliosphere~\citep{2012Ap&SS.339..223S}. Additionally, we note that there is no CR dataset which we can compare solar parameters to on day-long timescales.

In order to use this solar data to constrain the values of $\phi_{0}$ and $\phi_{1}$, and the functions $g(|B_{\rm tot}(t)|)$ and $f(\alpha (t))$, we compare the solar observables with the measurements of the CR proton spectrum taken between 1992 and 2007 by IMAX, BESS, \textit{AMS-01}, CAPRICE, and BESS Polar, and then continuously between July 2006 and January 2010 by \textit{PAMELA}. We note that our ability to constrain these parameters relies sensitively on the quantity of CR data available. At present, only \textit{PAMELA} and \textit{AMS-02} have acceptances large enough to detect variations in the CR proton spectrum that appear over month-long timescales. 

Since CRs with energy E$\sim$100~MeV typically take between $\sim$100 --- 300 days to travel from the heliopause to the Earth's location, depending on their charge and the solar activity in that period (with CRs traveling through the poles traveling faster) \cite{2012Ap&SS.339..223S}; we take different propagation time-scales for particles with $qA>0$ than $qA<0$. 
Throughout our analysis, we assume that CRs propagating from the poles ($qA>0$) take 3 months to arrive at the Earth, while CRs propagating through the heliospheric current sheet ($qA<0$) take between 3 and 12 months, depending on the average values of $|B_{\rm tot}|$ and $\alpha$. With this in mind, we average the values of $|B_{\rm tot}(t)|$ and $\alpha(t)$ used in our analysis over the 3-12 month periods preceding the time of data collection.

\begin{figure*}
\begin{centering}
\vspace{-1.2cm}
\includegraphics[width=7.6in,angle=0]{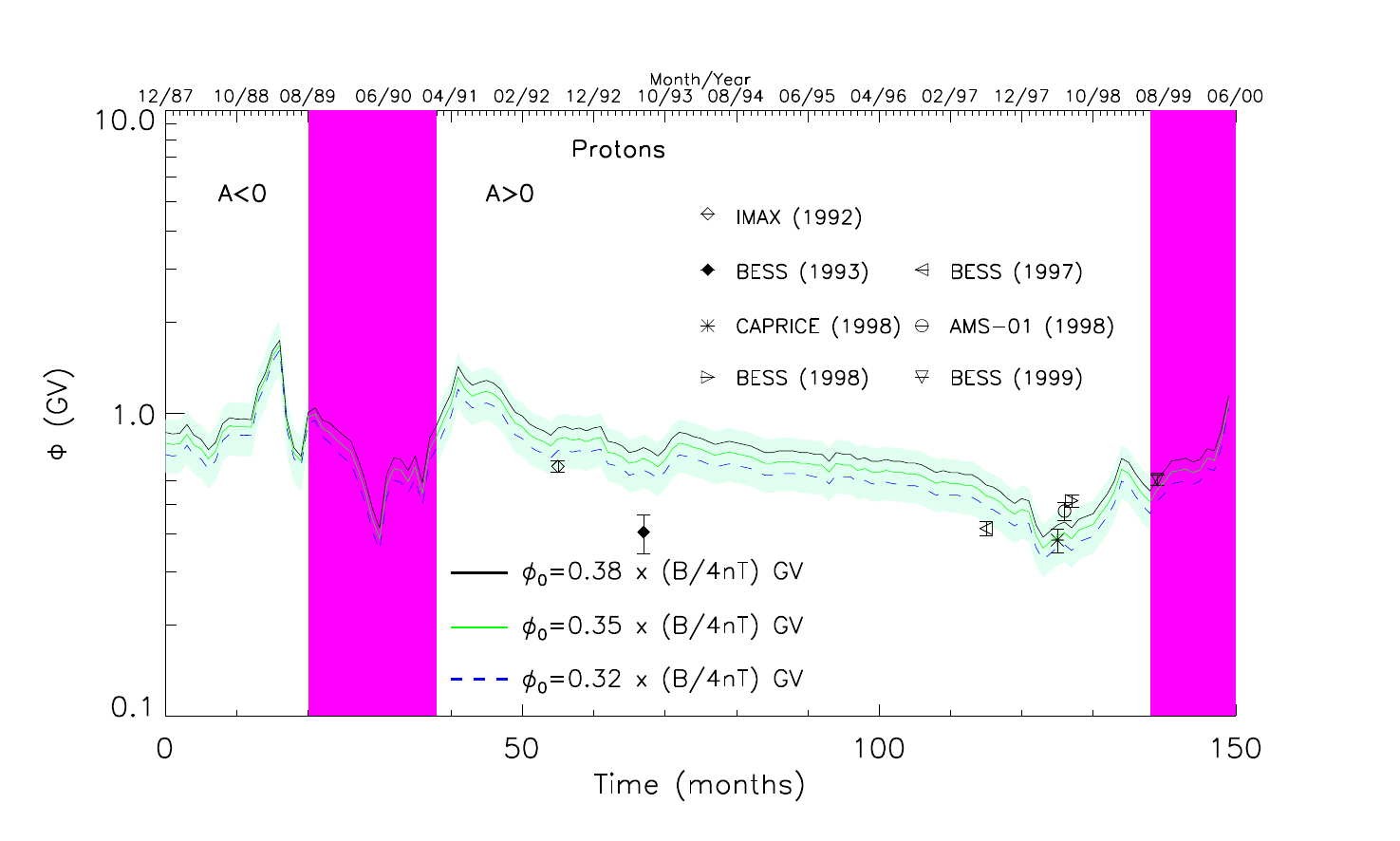} \\
\vspace{-1.2cm}
\includegraphics[width=7.6in,angle=0]{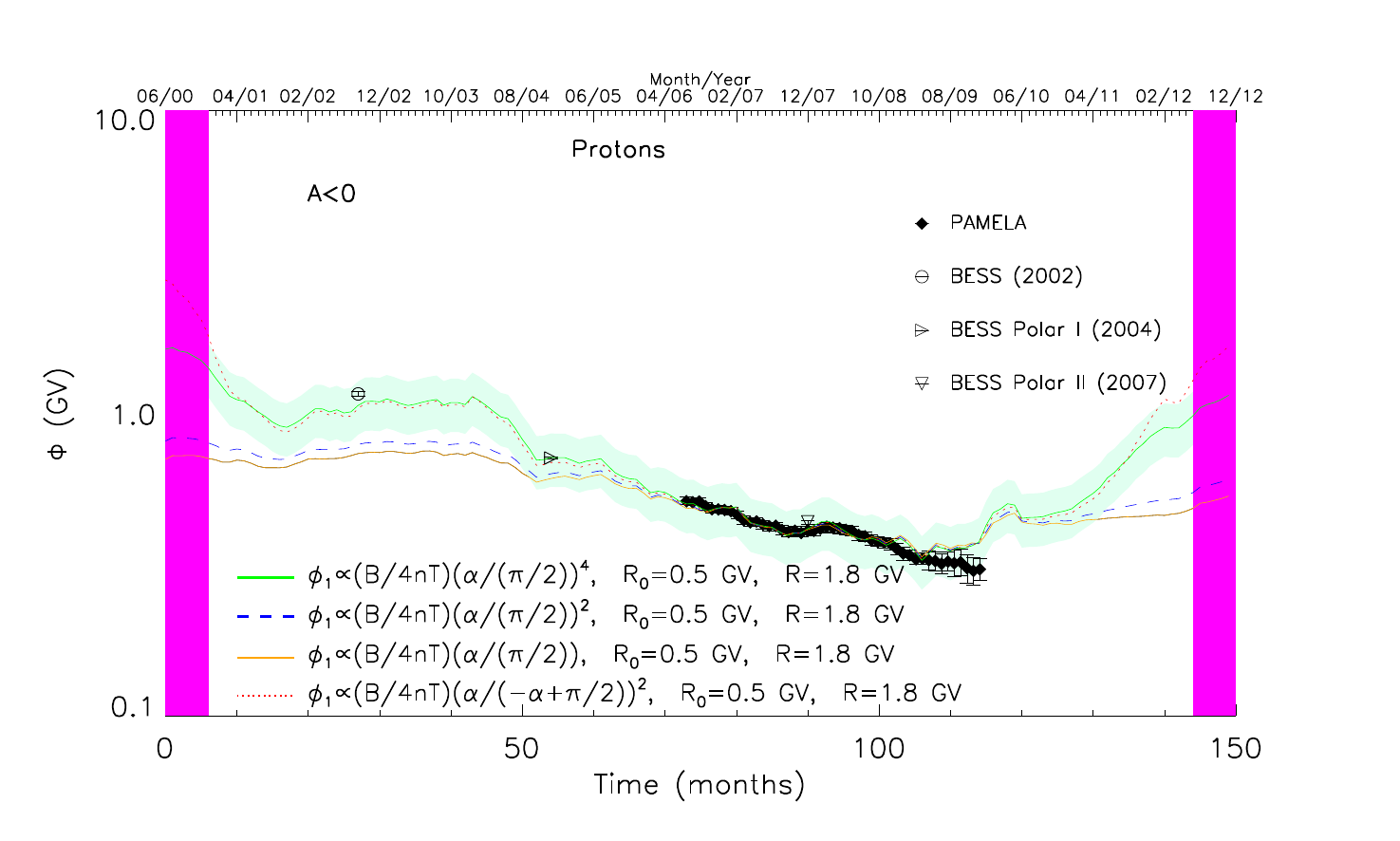}
\vspace{-1.0cm}
\end{centering}
%
\caption{Top: The calculated modulation potential in the time period between December 1987 (month 0) and June 2000 (month 150) for three different choices of the normalization parameter, $\phi_0$. Bottom: The calculated modulation potential in the time period between June 2000 (month 0) and December 2012 (month 150) for four different parameterizations of $\alpha(t)$. In each frame, the predictions are compared to the values (data points) of the pure force-field modulation potential inferred from fitting the CR proton spectrum as measured by IMAX~\cite{2000ApJ...533..281M}, BESS 93 \cite{2002ApJ...564..244W}, BESS 97 \cite{Shikaze:2006je}, CAPRICE 98 \cite{Boezio:2002ha}, \textit{AMS-01} \cite{Alcaraz:2000vp}, BESS 98 \cite{Sanuki:2000wh}, BESS 99 \cite{Asaoka:2001fv}, and \textit{PAMELA} \cite{Adriani:2013as}. In each frame, we show a band around the prediction of our default model, representing the estimated 20\% systematic uncertainty. The vertical purple bands span an 18 month period of time around each polarity flip, during which the configuration of the solar magnetic field changes drastically, limiting the ability of our model to make reliable predictions. These results were derived using the Galactic cosmic ray Model C, and the values of the modulation potential can vary by up to 10\% when the other models listed in Table~\ref{tab:fitResults} are instead used.}
\label{fig:ModPotvstime}
\end{figure*}

We begin by analyzing the data taken during the $A>0$ period between 1990 and 2000. For CR protons during this period of positive polarity, the Heaviside function in Equation~\ref{eq:main} is zero, thus allowing us to neglect any dependance on the tilt angle or rigidity, focusing instead on the parameter $\phi_0$ and the function $g(|B_{\rm tot}|)$. In the top frame of Figure~\ref{fig:ModPotvstime}, we show the predicted time dependence of the modulation potential for CR protons with arbitrary rigidity for three different values of $\phi_{0}$, assuming $g(|B_{\rm tot}(t)|) \propto |B_{\rm tot}(t)|$. We compare this prediction with the CR proton fluxes observed by IMAX~\cite{2000ApJ...533..281M}, BESS 93 \cite{2002ApJ...564..244W}, BESS 97 \cite{Shikaze:2006je}, CAPRICE 98 \cite{Boezio:2002ha}, \textit{AMS-01} \cite{Alcaraz:2000vp}, BESS 98 \cite{Sanuki:2000wh} and BESS 99 \cite{Asaoka:2001fv}. 

For each dataset, the error bars represent the range of the pure force-field modulation potential that provides a fit to the observed proton spectrum within 1$\sigma$ of the best fit value, starting with the (unmodulated) interstellar CR spectrum predicted by Model C ($\delta=0.4$), as described in Section~\ref{sec:ISM_Uncer}. These results (central values of the pure force-field modulation potential shown in Fig.~\ref{fig:ModPotvstime} with error-bars) varied by less than 10\% when the other Galactic CR models described in Section~\ref{sec:ISM_Uncer} were adopted instead.  We also allowed for some freedom in the CR proton normalization to account for systematic variations between different experiments.

Between May and July of 1998, CAPRICE, \textit{AMS-01} and BESS each independently measured the CR proton spectrum. While the results of these experiments are not mutually consistent at the 1$\sigma$ level, their combination allows us to strongly constrain the value of $\phi_{0}$ in the $qA>0$ regime, as the amplitude of $|B_{\rm tot}(t)|$ remained relatively constant over that period. We find a best fit value within 10\% of $\phi_{0}= 0.35$~GV for all five of our CR propagation models (for $g=|B_{\rm tot}|/4\,{\rm nT}$).

After estimating the value of $\phi_{0}$, we utilize IMAX 92, BESS 93, 97 and 99 data to constrain the dependance of the modulation potential on the amplitude of the HMF, $g(|B_{\rm tot}(t)|)$. Assuming that $g(|B_{\rm tot}(t)|)$ $\propto |B_{\rm tot}(t)|^{n}$, we find that only values in the range of $n \simeq 0-1$ are able to produce a reasonable fit to these datasets. This result, combined with the physical argument that the timescale for CR drift should be proportional to $B_{\rm tot}$ (see Equation~\ref{drift3}), leads us adopt $n=1$.\footnote{Although a value of $n=1$ is in some tension with the BESS 1993 dataset, 1993 was the first run of the BESS program, and we consider it possible that their systematic uncertainties were larger than reported. Indeed, we find that fits to the BESS data require the normalization of the interstellar CR proton spectrum to fall by 35\% during the period of the 1993 flight, and is suggestive instead of a systematic error in the experiment's acceptance.} 
%
%
We also note that the precise value of $n$ will soon be tested with high statistical precision by AMS-02 in the ongoing $A>0$ epoch.

Moving on, we next consider the modulation potential in the $qA<0$ regime.  We first note that the value of $\phi_0$ and functional form of $g \propto |B_{\rm tot}(t)|$ for the $qA>0$ case are also consistent with the data taken in periods of opposite polarity. Larger values of $\phi_{0}$ in the $qA<0$ era are ruled out by the \textit{PAMELA} data taken during the 2009 minimum, while smaller values of $\phi_{0}$ conflict with the full \textit{PAMELA} dataset. Similarly, the \textit{PAMELA} measurements rule out variations in the $\phi_0$ term with a dependence on the magnetic field intensity greater than $n=1$.

As an alternative, one could consider increasing the value of $\phi_1$ and simultaneously modifying the dependance on $|B_{\rm tot}|$ of the term proportional to $\phi_1$. We remind the reader, however, that a linear relationship between the modulation potential and $|B_{\rm tot}(t)|$ aligns well with theoretical expectations (see Equation~\ref{drift3}). Furthermore, we find that replacing $g(|B_{\rm tot}(t)|)$ in the $\phi_1$ term with a different functional form leads to disagreement with the variations of the modulation potential observed over month-long timescales, as well as with the average value measured by \textit{PAMELA} over the period spanning 2006-2008~\cite{Adriani:2011cu}.


In the bottom frame of Figure~\ref{fig:ModPotvstime}, we plot the modulation potential (evaluated at $R=1.8$ GV) as predicted using four different parameterizations for $f(\alpha (t))$, and for $\phi_{0} = 0.35$ GV and $g= |B_{\rm tot}(t)|/4\,{\rm nT}$. Although simulations have suggested a fairly weak correlation between the modulation potential and $\alpha$ ({\it ie.} $f = \alpha/(\alpha + \alpha_{0})$, with $\alpha_{0} \simeq 10^{\circ} -  30^{\circ}$) \cite{2012Ap&SS.339..223S, Potgieter:2013pdj}, we find that the combined \textit{PAMELA} and BESS data favors a significantly stronger dependance on $\alpha$. In particular, the best fits were found for $f = \alpha(t)^n$ with $n \simeq 3-6$, or $f \simeq [\alpha/(-\alpha + \alpha_0)]^2$ with $\alpha_0 \sim 90^{\circ}$. As our default model, we adopt $f = \alpha(t)^4$. We note that the preference for this parameterization relies strongly on the single data point from BESS 2002 (and to a lesser extent from BESS Polar I), leaving us less confident in this determination. Given the large tilt angles observed between 2010 and 2012, we anticipate that early data from \textit{AMS-02} will be very useful in further constraining the behavior of $f(\alpha (t))$.

Taking these result together, we arrive at the parameterization for the solar modulation potential presented earlier as Equation~\ref{eq:ModPot_Intro}, and repeated here:
\begin{eqnarray}
\Phi(R,t) = \phi_{0} \, \bigg( \frac{|B_{\rm tot}(t)|}{4\, {\rm nT}}\bigg) + \phi_{1} \, H(-qA(t))\, \bigg( \frac{|B_{\rm tot}(t)|}{4\,  {\rm nT}}\bigg) \, \bigg(\frac{1+(R/R_0)^2}{\beta (R/R_{0})^3}\bigg) \, \bigg( \frac{\alpha(t)}{\pi/2} \bigg)^{4},
\label{eq:ModPot_Final}
\end{eqnarray}
with $\phi_{0} = 0.35$ GV, $\phi_{1}=3.9$ GV and $R_0=0.5$ GV.\footnote{In fitting the free parameters of our model, we have required $R_{0} > 0.4$ GV, motivated by the results of earlier work that suggest $R_{0}  \sim O(1)$ GV~\cite{2000JGR...10527447B,Strauss:2012baa}. }

Regarding the uncertainties associated with the above expression, we have found variations at the level of up to 10\% between the five different galactic CR models described in Section~\ref{sec:ISM_Uncer}. We note that the formal 1 $\sigma$ errors on the modulation potential from fitting the CR spectra are significantly smaller than the systematic uncertainties stemming from our ignorance of the true interstellar CR proton spectrum. For example, in Model C we obtain a force-field approximation value of $\Phi$~=~$1.157^{+0.020}_{-0.018}$ GV, while for Model E we obtain $\Phi$~=~$1.110^{+0.022}_{-0.017}$ GV. Secondly, we note that if we include an additional 10\% systematic error, the predicted values of $\Phi(R,t)$ yield a $\chi^{2}/dof$ of less than 1.0 when fit to the time-dependent \textit{PAMELA} dataset (see Figure~\ref{fig:ModPotvstime}). Taken together, we estimate that the expression for the modulation potential in Equation~\ref{eq:ModPot_Final} has a total systematic uncertainty of approximately 20\%. The exception to this statement is that the uncertainties are likely to be larger during periods leading up to and following a polarity flip. Since each polarity flip lasts about half a year~\cite{Polarsite} and the CRs take anywhere between $\simeq 100$ days to a year to arrive at Earth, we do not expect our model to yield accurate predictions for a period of approximately 18 months surrounding each change in polarity. These periods of time are depicted as vertical purple bands in Figure~\ref{fig:ModPotvstime}. 

\begin{figure*}
\begin{centering}
\vspace{-1.2cm}
\includegraphics[width=7.6in,angle=0]{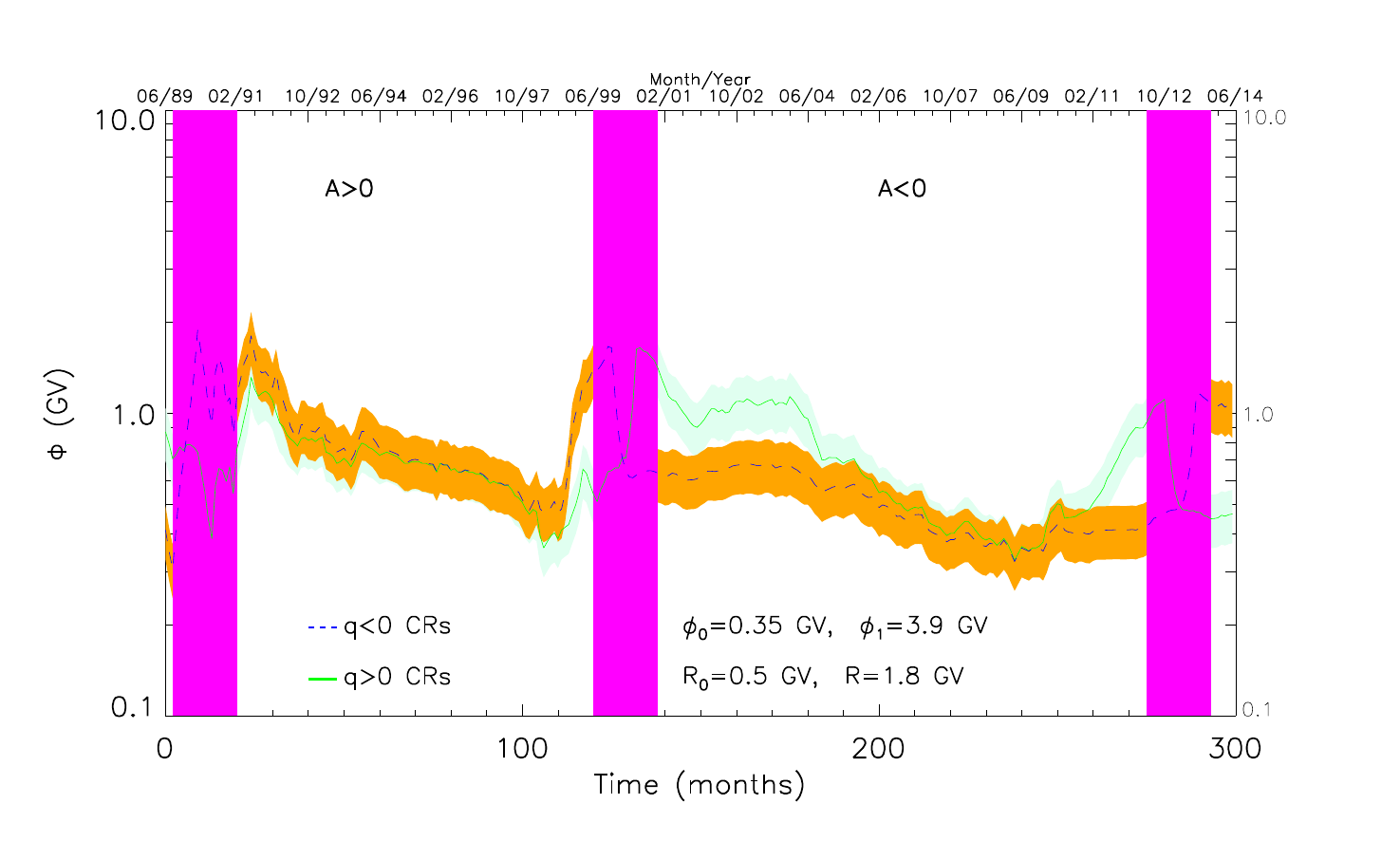} \\
\vspace{-1.2cm}
\includegraphics[width=7.6in,angle=0]{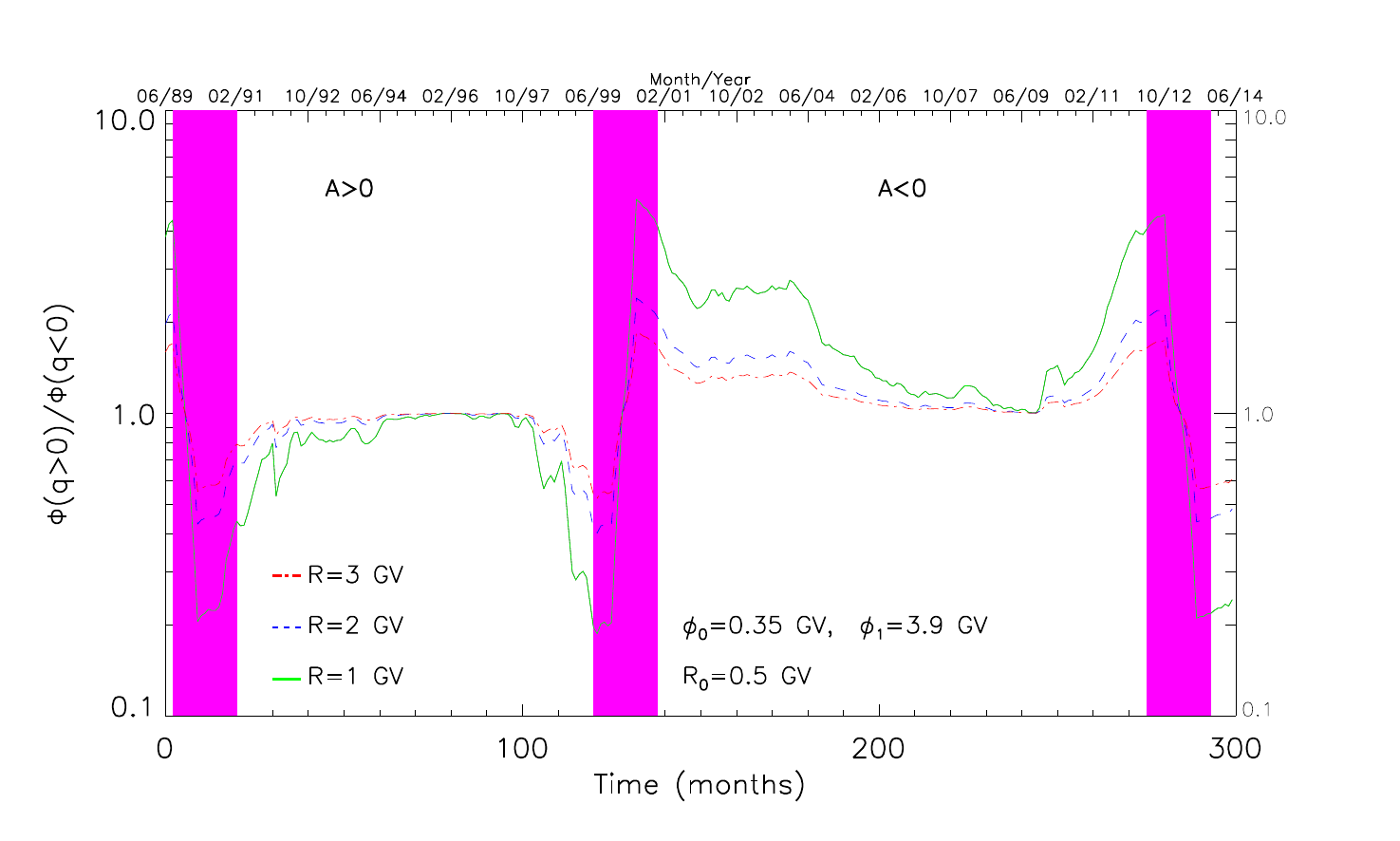}
\vspace{-1.0cm}
%
\end{centering}
\caption{Top: The modulation potential predicted by Equation~\ref{eq:ModPot_Final} in the time period between June 1989 (month 0) and June 2014 (month 300), for both positively and negatively charged particles. The shaded bands around these curves represent the estimated 20\% systematic uncertainty. Bottom: The ratio of the modulation potentials for positively charged and negatively charged cosmic rays, over the same time period, for three values of rigidity. Again, the vertical purple bands span an 18 month period of time around each polarity flip, during which the configuration of the solar magnetic field changes drastically, limiting the ability of our model to make reliable predictions.}
\label{fig:ModPotBothCharges}
\end{figure*}

In the top frame of Figure~\ref{fig:ModPotBothCharges}, we plot the value of the modulation potential predicted by our model as a function of time, for both positively and negatively charged CRs, evaluated at a rigidity of $R=1.8$ GV. The bands around each curve reflect the estimated 20\% systematic uncertainties described in the previous paragraph. Among other features, this figure demonstrates that CRs in eras with $qA<0$ experience more significant variations of the modulation potential with time. This is in agreement with observations of CR protons, helium nuclei, and electrons made by the \textit{Ulysses} experiment~\cite{2002JGRA..107.1274H, 2006SSRv..127..117H} (for additional discussion, see Ref.~\cite{Potgieter:2013pdj}). In the bottom frame of the same figure, we plot the predicted ratio of the modulation potentials for positively and negatively charged CRs, for three values of the rigidity. While this ratio is often found to be near unity, significant charge-dependent modulation is predicted over some periods of time, and in particular for low-rigidity CRs.

\begin{table*}[t]
\begin{tabular}{cccccccccc}
\hline
Era & Exper. & $|B_{\rm tot}|$ (nT) & $\alpha$ (degrees) & $\Phi^{(q>0)}_{R=1GV}$ & $\Phi^{(q>0)}_{R=2GV}$ & $\Phi^{(q>0)}_{R=3GV}$ & $\Phi^{(q<0)}_{R=1GV}$ & $\Phi^{(q<0)}_{R=2GV}$ & $\Phi^{(q<0)}_{R=3GV}$ \\
\hline \hline
07/92 & IMAX & 8.9 & 32.1 & 0.78 & 0.78 & 0.78 & 0.90 (0.89) & 0.82 (0.82) & 0.80 (0.80) \\
07/93 & BESS & 7.9 & 35.4 & 0.69 & 0.69 & 0.69 & 0.85 (0.80) & 0.75 (0.73) & 0.72 (0.71) \\
07/97 & BESS & 6.4 & 22.6 & 0.56 & 0.56 & 0.56 & 0.58 (0.62) & 0.57 (0.58) & 0.56 (0.57) \\
05/98 & CAPRICE & 4.3 & 46.3 & 0.38 & 0.38 & 0.38 & 0.63 (0.45) & 0.46 (0.40) & 0.43 (0.39) \\
06/98 & \textit{AMS-01} & 4.5 & 45.2 & 0.39 & 0.39 & 0.39 & 0.63 (0.47) & 0.48 (0.42) & 0.44 (0.41) \\
07/98 & BESS & 4.6 & 46.6 & 0.40 & 0.40 & 0.40 & 0.68 (0.49) & 0.50 (0.43) & 0.46 (0.42) \\
07/99 & BESS & 5.8 & 73.9 & 0.51 & 0.51 & 0.51 & 2.71 (0.67) & 1.26 (0.56) & 0.97 (0.54) \\
08/02 & BESS & 7.6 & 55.1 & 1.54 (0.83) & 0.96 (0.72) & 0.85 (0.70) & 0.66 & 0.66 & 0.66 \\
12/04 & BESS Polar I & 6.4 & 46.5 & 0.95 (0.68) & 0.69 (0.60) & 0.64 (0.59) & 0.56 & 0.56 & 0.56 \\
07-12/06 & \textit{PAMELA} & 5.2 & 34.2 & 0.54 (0.52) & 0.48 (0.48) & 0.47 (0.47) & 0.45 & 0.45 & 0.45 \\
01-06/07 & \textit{PAMELA} & 4.9 & 32.1 & 0.49 (0.49) & 0.45 (0.45) & 0.44 (0.44) & 0.43 & 0.43 & 0.43 \\
07-12/07 & \textit{PAMELA} & 4.4 & 31.1 & 0.44 (0.44) & 0.40 (0.40) & 0.40 (0.40) & 0.39 & 0.39 & 0.39 \\
12/07 & BESS Polar II & 4.5 & 32.5 & 0.45 (0.44) & 0.41 (0.41) & 0.40 (0.40) & 0.39 & 0.39 & 0.39 \\
01-06/08 & \textit{PAMELA} & 4.5 & 34.7 & 0.47 (0.45) & 0.42 (0.41) & 0.41 (0.41) & 0.39 & 0.39 & 0.39 \\
07-12/08 & \textit{PAMELA} & 4.2 & 28.8 & 0.40 (0.41) & 0.38 (0.38) & 0.37 (0.38) & 0.37 & 0.37 & 0.37 \\
01-06/09 & \textit{PAMELA} & 4.0 & 21.5 & 0.36 (0.38) & 0.36 (0.36) & 0.35 (0.36) & 0.35 & 0.35 & 0.35 \\
07-12/09 & \textit{PAMELA} & 4.1 & 18.7 & 0.36 (0.39) & 0.36 (0.37) & 0.36 (0.36) & 0.36 & 0.36 & 0.36 \\
01-06/10 & \textit{PAMELA} & 4.7 & 39.7 & 0.56 (0.48) & 0.46 (0.44) & 0.44 (0.43) & 0.41 & 0.41 & 0.41 \\
07-12/10 & \textit{PAMELA} & 4.6 & 39.9 & 0.55 (0.47) & 0.45 (0.43) & 0.43 (0.42) & 0.40 & 0.40 & 0.40 \\
01-06/11 & \textit{PAMELA} & 4.7 & 48.3 & 0.73 (0.50) & 0.52 (0.44) & 0.48 (0.43) & 0.41 & 0.41 & 0.41 \\
07-12/11 & \textit{AMS-02}/\textit{PAMELA} & 4.7 & 60.5 & 1.21 (0.52) & 0.69 (0.45) & 0.58 (0.43) & 0.41 & 0.41 & 0.41 \\
01-06/12 & \textit{AMS-02}/\textit{PAMELA} & 4.8 & 67.2 & 1.66 (0.54) & 0.85 (0.46) & 0.68 (0.45) & 0.42 & 0.42 & 0.42 \\
01-06/14 & \textit{AMS-02} & 5.3 & 67.3 & 0.46 & 0.46 & 0.46 & 1.83 (0.60) & 0.92 (0.51) & 0.75 (0.49) \\
07-12/14 & \textit{AMS-02} & 5.6 & 62.0 & 0.49 & 0.49 & 0.49 & 1.54 (0.62) & 0.85 (0.54) & 0.71 (0.52) \\
01-06/15 & \textit{AMS-02} & 6.6 & 56.6 & 0.58 & 0.58 & 0.58 & 1.44 (0.72) & 0.87 (0.63) & 0.76 (0.61) \\
07-12/15 & \textit{AMS-02} & 7.0 & 51.5 & 0.61 & 0.61 & 0.61 & 1.24 (0.75) & 0.83 (0.66) & 0.74 (0.64) \\
01-06/16 & \textit{AMS-02} & 6.7 & 48.8 & 0.59 & 0.59 & 0.59 & 1.07 (0.71) & 0.75 (0.63) & 0.69 (0.61) \\
\hline \hline
\end{tabular}
\caption{The total modulation potential $\Phi$ (in GV) from Equation~\ref{eq:ModPot_Final}, for different eras probed by different experiments. We give the averaged values of $|B_{\rm tot}|$ and $\alpha$ relevant for each time period. Given the rigidity dependence of $\Phi$ for $qA<0$, we provide its value for three different rigidities. The values of the modulation potential given in parentheses were derived using $f \propto \alpha$, which is disfavored by BESS 2002 and BESS Polar I, rather than our default parameterization of $f \propto \alpha^4$. See text for details.}
\label{tab:Phi_vs_Experiment}
\end{table*}

In Table~\ref{tab:Phi_vs_Experiment}, we provide the values of the HMF and tilt angle of the heliospheric current sheet as measured over the time periods that measurements were carried out by the IMAX, BESS, CAPRICE, AMS-01, BESS Polar, PAMELA, and AMS-02 experiments. These time-dependent quantities are provided here for convenient use in Equation~\ref{eq:ModPot_Final}. We also include in this table the modulation potential predicted for each of these time periods, for positively and negatively charged CRs, and for three values of their rigidity. The quantities in parentheses represent the values of the modulation potential as derived using $f \propto \alpha$, rather than our default choice of $f \propto \alpha^4$. Although this linear relationship is significantly disfavored by the BESS 2002 and BESS Polar I measurements, we include these results in acknowledgement of the more significant uncertainties associated with the modulation potential during periods with $qA<0$. Forthcoming data from \textit{AMS-02} is expected to much more tightly constrain the relationship between the modulation potential and $\alpha(t)$.

\section{Summary and Conclusions}
\label{sec:Conclusions}

In recent years, the effects of solar modulation have often limited our ability to constrain models for the production and propagation of cosmic rays throughout the Milky Way. In many cosmic-ray studies, solar effects are treated by applying a simple force-field modulation potential, with a value that is chosen to provide the best fit to the cosmic-ray dataset under consideration. The value of the modulation potential is often effectively degenerate with parameters associated with other physical phenomena, such as convection and diffusive reacceleration.

In this study, we have made use of time-dependent measurements of the magnitude and polarity of the heliospheric magnetic field and the tilt angle of the heliospheric current sheet, in combination with a variety of measurements of the cosmic-ray spectrum at Earth, and outside of the influence of the solar wind, as recently measured by \textit{Voyager 1}. Through this approach, we have constrained the relationship between the modulation potential and solar observables, allowing us to produce an analytic expression for the modulation potential that is dependent on time, charge and rigidity (see Equation~\ref{eq:ModPot_Final}). Instead of treating the modulation potential for a given measurement as a nuisance parameter, one can use this equation to calculate the modulation potential for a given charge and rigidity at a given polarity $A$ era and including publicly available information for the magnetic field amplitude and heliospheric current sheet tilt angle \cite{ACESite, WSOSite}.

We have constrained the functional form and free parameters in our analytic expression using the data taken over the past 24 years by a variety of cosmic ray experiments, including IMAX, BESS, CAPRICE, BESS Polar, \textit{AMS-01}, \textit{PAMELA}, and \textit{Voyager 1}. Data from \textit{AMS-02} is expected to significantly improve our ability to constrain the precise form of this model. Assuming limited systematic uncertainties, proton data from \textit{AMS-02} taken up to June 2012 ($A<0$) and from January 2014 ($A>0$) is expected to be sensitive to the changes in the modulation potential at the level of a few percent, allowing us to tightly constrain the dependence of this quantity on the value of the tilt angle of the heliospheric current sheet and cosmic-ray rigidity. CR electrons and antiprotons suffer from additional astrophysical uncertainties related mainly to their energy losses ($e^{-}$) and production rate ($\bar{p}$) and thus are suboptimal compared to the CR protons for such a study. CR positrons suffer from both uncertainties in their energy losses (as $e^{-}$) and from uncertainties related to their sources.

One of the key science goals of \textit{AMS-02} and other cosmic-ray experiments is to search for the antimatter cosmic rays that are predicted to be produced in the annihilations of weak-scale dark matter particles.  Measurements of the cosmic-ray antiproton spectrum have already been used to place strong constraints on the dark matter annihilation cross section~\citep{Evoli:2011id,Fornengo:2013xda,Hooper:2014ysa,Cirelli:2014lwa,Cholis:2010xb,Cirelli:2008pk,Donato:2008jk,Garny:2011cj,Chu:2012qy,Belanger:2012ta,Cirelli:2013hv,Bringmann:2014lpa}, comparably stringent to those from gamma-ray observations of dwarf galaxies~\cite{Ackermann:2015zua,Geringer-Sameth:2014qqa} and the Galactic Center~\cite{Hooper:2012sr}. Similarly, cosmic-ray positron measurements have been used to place strong constraints on the dark matter annihilation cross section to charged leptons~\cite{Bergstrom:2013jra}. Our ability to constrain dark matter annihilation with cosmic-ray data is currently limited in large part by systematic uncertainties, including those associated with the anti-proton production cross section~\cite{Kappl:2014hha,Moskalenko:2001qm,diMauro:2014zea}, and with the modeling of cosmic-ray injection and propagation through the Galaxy and Solar System~\cite{Evoli:2011id,Fornengo:2013xda,Hooper:2014ysa,Cirelli:2014lwa}. We expect that the model for solar modulation presented here (and its future refinements) will be helpful in reducing these systematic uncertainties, and enabling cosmic-ray experiments to increase their sensitivity to dark matter annihilation products.




\bigskip  

{\it Note added}: After the completion of this work, we became aware of a related parallel study by Corti, Bindi, Consolandi and Whitman \cite{Corti:2015bqi}. Using similar data sets they have independently reached the conclusion for a  needed deviation from the force field approximation to account for  the rigidity dependence of the solar modulation of CRs. In addition an other related study by Kappl \cite{Kappl:2015hxv} were a publicly available code \emph{SOLARPROP} was released
 after the completion of this work. \cite{Kappl:2015hxv} is in agreement with our results that a simple 
one parameter time dependent description of solar modulation is insufficient to describe the data. 
                  
\bigskip                  
                  
{\it Acknowledgements}: We would like to thank V. Bindi, M. Boezio, C. Corti, A. Ibarra, T. Larsen, C. Weniger and S. Wild for valuable discussions. IC is supported by NASA grant NNX15AB18G and would like to thank the Korea Institute for Advanced Study for their hospitality provided during the completion of this work. DH is supported by the US Department of Energy under contract DE-FG02-13ER41958. Fermilab is operated by Fermi Research Alliance, LLC, under Contract No. DE- AC02-07CH11359 with the US Department of Energy. TL is supported by the National Aeronautics and Space Administration through Einstein Postdoctoral Fellowship Award No. PF3-140110.
                  
\bibliography{SolarModulation}
\bibliographystyle{apsrev}

\begin{appendix}

\section{Comparison With Other Cosmic-Ray Measurements}
\label{app2}

\begin{figure*}
\begin{centering}
\includegraphics[width=5.40in,angle=0]{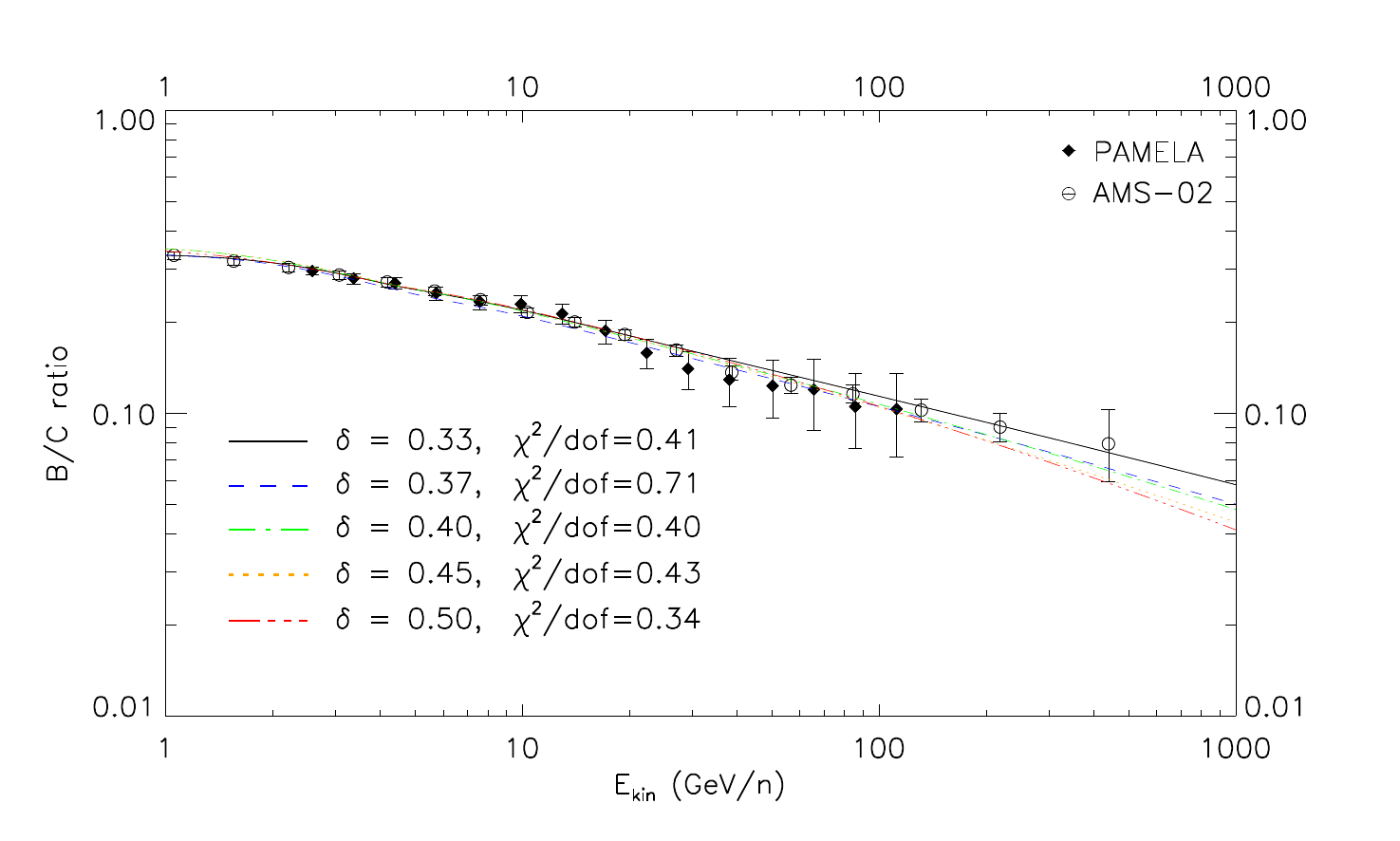}\\
\includegraphics[width=5.40in,angle=0]{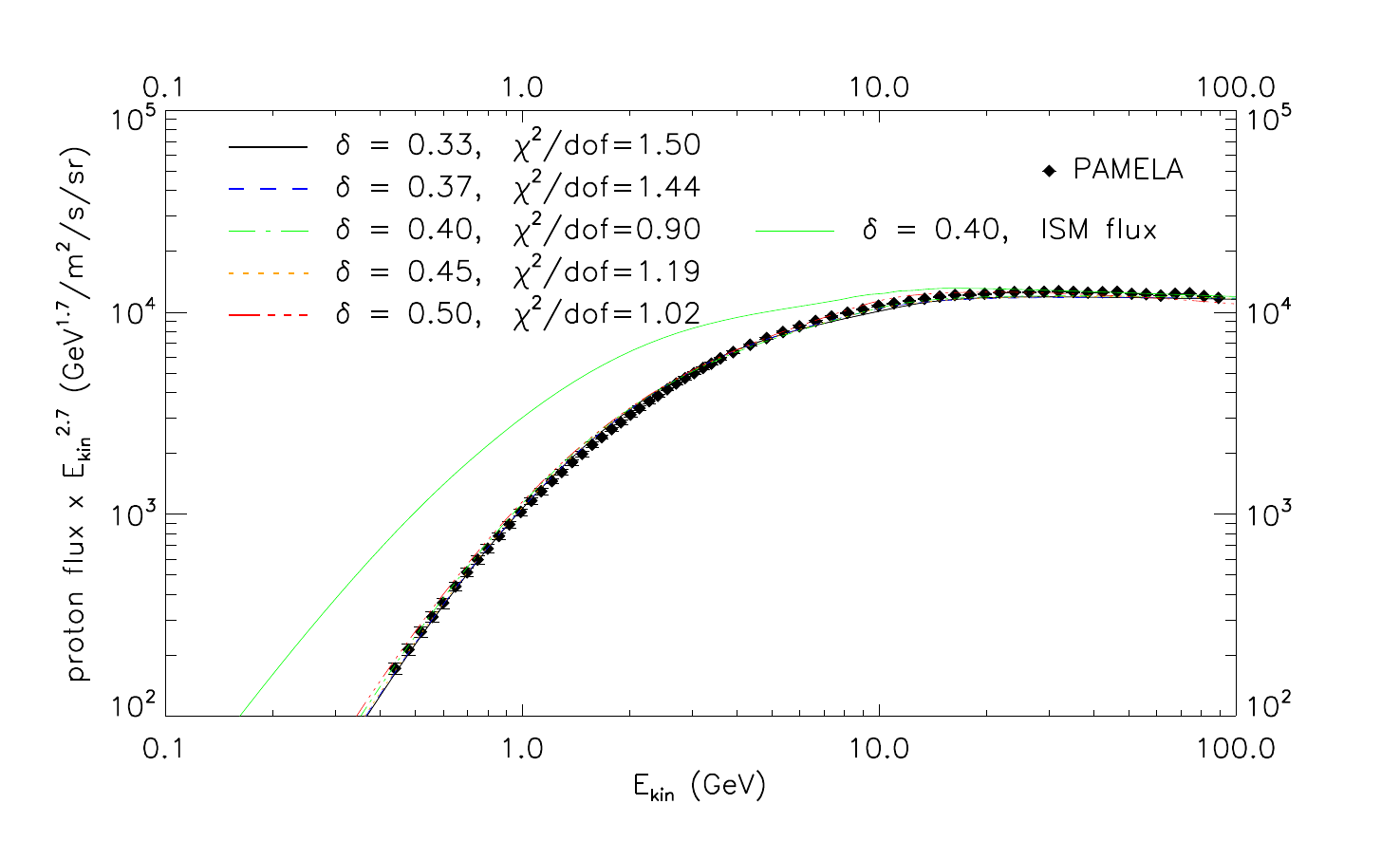}
\end{centering}
\caption{Top: The CR boron-to-carbon ratio predicted for the various Galactic cosmic-ray models given in Table~\ref{tab:fitResults}. Bottom: A comparison between the cosmic-ray proton spectrum for the same set of models and the \textit{PAMELA} data. In each frame, we have applied the model of solar modulation presented in this paper. For protons and for model C ($\delta = 0.40$) with solid green line we show the unmodulated ISM flux for comparison.}
\label{fig:CRfit}
\end{figure*}

\begin{figure*}
\begin{centering}
\includegraphics[width=5.40in,angle=0]{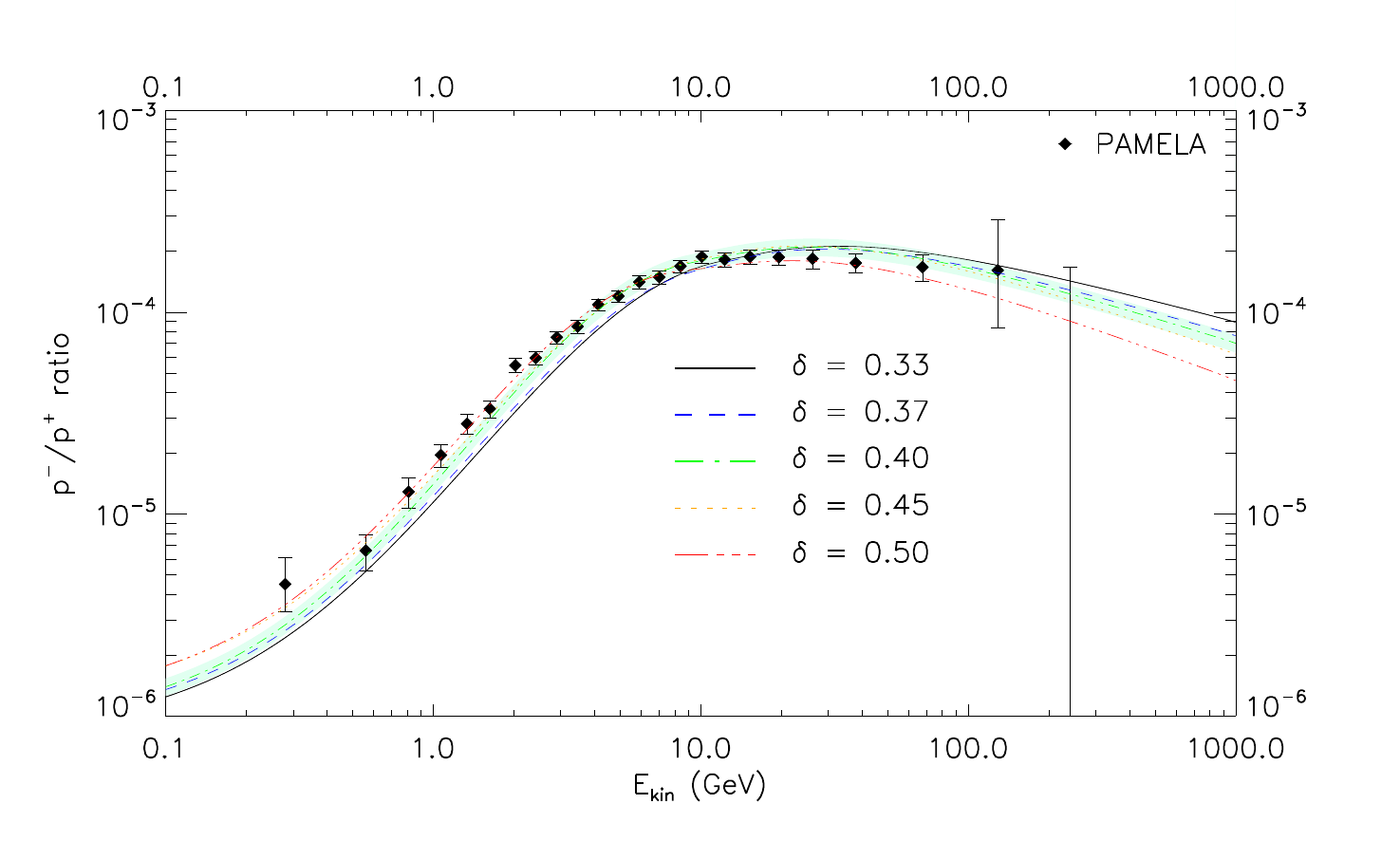}\\
\end{centering}
\caption{The predicted antiproton-to-proton ratio for the various Galactic cosmic-ray models given in Table~\ref{tab:fitResults}, including the application of the model of solar modulation presented in this paper. The light green band centered around model C ($\delta = 0.40$) reflects the $\pm 10 \%$ uncertainty associated with the local gas density and the antiproton production cross section.}
\label{fig:CRpredicted}
\end{figure*}

In Figure~\ref{fig:CRfit}, we plot the boron-to-carbon ratio as measured by \textit{PAMELA} and \textit{AMS-02}, and the proton spectrum as measured by \textit{PAMELA} \footnote{An alternative indirect measure of the CR proton spectrum is through gamma-ray data \cite{Strong:2015yva, Casandjian:2015hja}, since the $\pi^{0}$ emission is the largest galactic diffuse component.}, and compare this to the predictions from each of the five Galactic CR models described in Section~\ref{sec:ISM_Uncer}, after experiencing solar modulation as described by Equation~\ref{eq:ModPot_Final}.\footnote{Note that in this figure, we have used the modulation potential as predicted during the period of \textit{PAMELA}'s measurement. As solar modulation only slightly impacts the boron-to-carbon ratio above 1 GeV/n, we have chosen to also include the \textit{AMS-02} data in this figure.} We find that each of these models yields an acceptable fit to this data: $\chi^{2}/dof \simeq 0.33-0.71$ for boron-to-carbon and $\chi^{2}/dof \simeq 0.9-1.5$ for the proton spectrum. The best fit was found using Model C, while Models A and B provided the worst fits.

Finally, in Figure~\ref{fig:CRpredicted} we plot the predicted antiproton-to-proton ratio, including the effects of solar modulation, for each of the five Galactic CR models described in Section~\ref{sec:ISM_Uncer}. We also plot a $\pm 10 \%$ band centered around model C ($\delta = 0.40$), to indicate the minimal uncertainties associated with the local gas density and the antiproton production cross section. We intend to further discuss the implications of the $\bar{p}/p$ spectrum in a future study. 

For the \textit{AMS-02} era that means that only the measurements up to June 2012 ($A<0$) and from January 2014 ($A>0$) are going to be useful to constrain further the modulation potential form. CR electrons and antiprotons suffer from additional astrophysical uncertainties related mainly to their energy losses ($e^{-}$) and production rate ($\bar{p}$) and thus are suboptimal compared to the CR protons for such a study.

\end{appendix}

\end{document}